%% file: svd2.tex
\documentclass[prl,floatfix,superscriptaddress,nofootinbib,shownopacs,shownokeys,preprintnumbers,twocolumn]{revtex4}

\usepackage[dvips]{epsfig}
\usepackage{graphicx}

\begin{document}

\title{The status and physics program of the Spectrometer with Vertex Detector.\footnote{Sumbitted to The Sixth International Workshop "Very High Multiplicity Physics", Dubna, Russia, 16-17 April, 2005}}

\include{authors}

\date{\today.}

\begin{abstract}
The brief history, physics program and the current status of the SVD-2 detector is presented. The future plans for the experiments with upgraded SVD-2M setup is discussed.
\end{abstract}

\maketitle

\section {The E-161 experiment.}

\begin{figure}[h]
\centering
\includegraphics[width=0.5\textwidth]{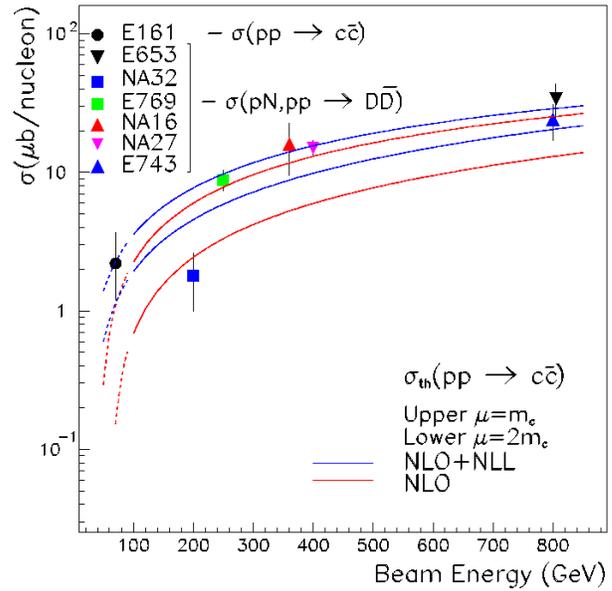}
\caption{The cross-section of $c \bar{c}$ pair production for proton-nucleon interactions.}
\label{charm}
\end{figure}

The Spectrometer with Vertex Detector (SVD) was originally constructed by the collaboration of IHEP, JINR, NPI MSU and Tbilisi IHEP institutes for the E-161 experiment at the Serpukhov Accelerator U70. The experiment was aimed at the study of charmed particle production in $pp$-interactions at $70~GeV$.
In the first stage of experiment (SVD-1) the rapid cycling liquid hydrogen bubble chamber (HBC) being the precision vertex detector and hydrogen target\cite{hbc} was used in combination with a wide aperture magnetic spectrometer with multiwire proportional chambers. HBC provided a track impact parameter measurement with a high precision of $3~\mu m$.
About 140000 $pp$-interactions at 70 GeV were selected in the fiducial bubble chamber volume, corresponding to a sensitivity of the experiment of 5~events/$\mu b$. After analysis of the combined bubble chamber and spectrometer information 3 decays of charged particles and 2 decays of neutral particles were selected as the charmed meson candidates. As a result an estimation of the total charm production cross section in $pp$-interactions at $70~GeV$ for $-1 < x_F < 1$ was obtained as
$1.6^{+1.1}_{-0.7}(stat.)\pm 0.3(syst.)~\mu b$\cite{svd1}. This result together with the other experimental data at high energies\cite{charm_e} and theoretical predictions from \cite{charm_t} is presented in Fig. \ref{charm}. This value does not contradict the "beam dump" experiment data at 70 GeV\cite{beamdump} and recent theoretical predictions\cite{ko}.

\section {The SVD-2 setup.}

In 1999-2002 the SVD detector has been seriously upgraded in order to increase the rate of event collection and to enhance its physics capabilities. The upgrade included the high resolution silicon microstrip vertex detector (instead of HBC), gamma-quanta detector and gas Cherenkov counter for particle identification.

The SVD-2 layout is presented on Fig. \ref{setup}. The SVD-2 setup includes the following basic components:

\begin{center}
\begin{figure}[ht]
\vspace{70mm}
{\includegraphics{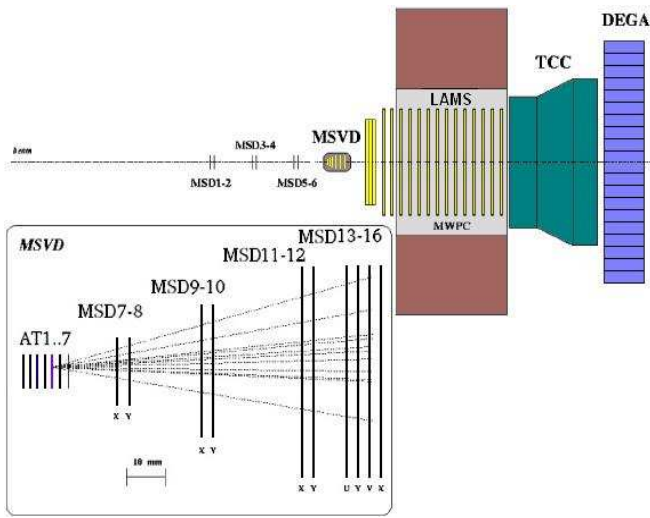}}
\caption{SVD-2 layout.}
\label{setup}
\end{figure}
\end{center}

\begin{enumerate}
\item The high-presicion microstrip vertex detector(MSVD). MSVD has the following elements:
\begin{itemize}
\item MSBT(microstrip beam telescope) - 3 XY pairs of microstrip Si-detectors (MSD1-6 each having 128 strips with 50$~\mu m$-pitch ), measuring XY-coordinates.
\item AT(active target) - 5 Si-detectors with a thickness of $300~\mu m$,
Pb-foil ($220~\mu m$) and C-target ($500~\mu m$). The distance between all the detectors is $4\ mm$.
\item MSVD(microstrip vertex detector) - 3 XY pairs of microstrip Si-detectors - MSD7-8 of 640 strips/$25~\mu m$-pitch, MSD9-10 (640/50 $~\mu m$), MSD11-12 (1024/50$~\mu m$) and UYVX quadruplet, MSD13-16 (1024/50$~\mu m$).
\end{itemize}
\item Large aperture magnetic spectrometer(LAMS) has the following main elements:
\begin{itemize}
\item Electromagnet MC-7A with aperture of $1.8\times1.3\ m^2$ and homogenous field of $1.18\ T$ over $3\ m$ long region.
\item A two sets of MWPC(multiwire proportional chambers). First one consists of 1 UYV triplet (sensitive area $1.0\times1.0\ m^2$ and $2~mm$ interwire distances) and placed before magnet, the second one (between magnet poles) consists of 5 UYV triplets with 2mm interwire distances and sensitive area of $1.0\times1.5\ m^2$. The part of these chambers are placed in the weak magnetic field.
\end{itemize}
\item The multicell threshold Cherenkov counter(TCC). The counter is designed for charged particles identification, it has an entrance aperture of $177\times 130\ cm^2$. Four rows with 8 spherical mirrors in each one are arranged on the rear wall of the counter. The threshold momentum for pions and protons is $4~GeV/c$ and $21~GeV/c$ respectively. When filled with freon at the atmospheric pressure and the temperature of $20 ^\circ C$, the counter provides the identification of pions in momentum range $4-21\ GeV/c$ with $70\%$ efficiency.
\item The gamma quanta detector (DEGA). DEGA consists of 1536 Cherenkov full absorption lead glass counters with the transverse dimensions of $38\times 38\ mm^2$ and the length of $505\ mm$. The total sensitive area of the detector is $1.8\times 1.2\ m^2$. DEGA provides a $\gamma$ registration with energies from $50\ MeV$ to $20\ GeV$ with the position resolution of $2-3\ mm$.
\end{enumerate}

The SVD-2 trigger system provides the trigger signal, based on data from scintillation counters before active target, from scintillation hodoscope, situated after TCC and from the active target strips. More detailed description of the trigger electronics and the data acquisition system may be found elsewhere\cite{svdtrig}.

\begin{center}
\begin{figure}[ht]
\vspace{55mm}
{\includegraphics{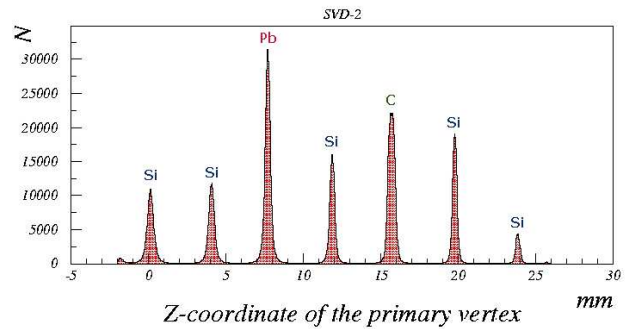}}
\caption{The reconstructed Z-coordinate of the primary vertex.}
\label{zvert}
\end{figure}
\end{center}

Data taking during the first physics run (April, 2002) was performed in the proton beam of IHEP accelerator with energy $E_p = 70\ GeV$ and intensity $I \approx (5 \div 6) \cdot 10^5$ p/cycle. The total statistics of $5\cdot10^7$ inelastic events was obtained. The sensivity of this experiment for inelastic $pN$-interactions taking in account the triggering efficiency was $1.6~nb^{-1}$. The primary vertex resolution was estimated as $70-120~\mu m$ for Z-coordinate (Fig. \ref{zvert}) and $8-12~\mu m$ for X(Y)-coordinates. For the two-tracks secondary vertices  ($K^0_s, \Lambda$) these values were $250~\mu m$ and $15~\mu m$ respectively.
The impact parameter resolution for $3-5\ GeV$ momentum tracks is
about  $12~\mu m$. The angular acceptance of the vertex detector
is averaged to $\pm 250\ mrad$.

\begin{figure}[ht]
\vspace{95mm}
{\includegraphics{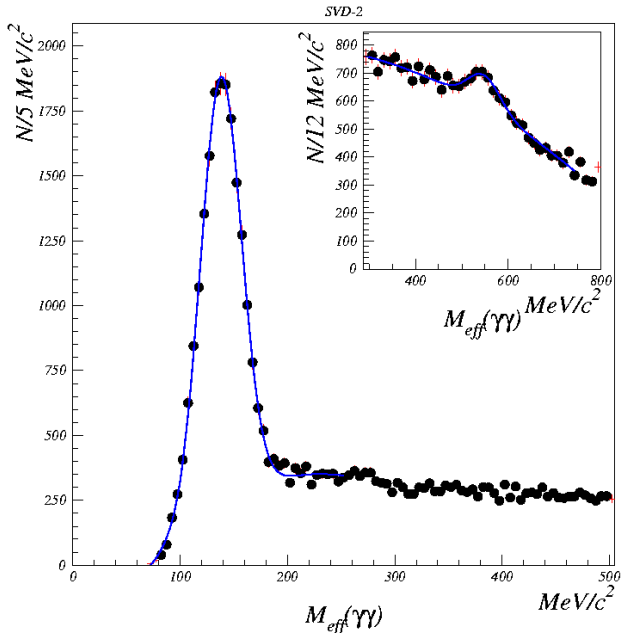}}
\caption{The $(\gamma \gamma)$ invariant mass spectrum. The inset shows the $\eta$-meson mass region.}
\label{pi0}
\end{figure}

The SVD-2 setup permits to obtain the high effective
mass resolution of $\sigma = 4.4~MeV/c^2$ for $K^0_s$ and $1.6~MeV/c^2$ for $\Lambda^0$ masses. The
momentum resolution for the track with 15 measured hits is
$(0.5\div1.0)\%$ in the $(4\div20)\ GeV/c$ momentum range. The
angular measurement error was estimated to be
$0.2\div0.3\ mrad$. The angular acceptance of the spectrometer
is averaged to $\pm200\ mrad$ for horizontal and $\pm150\ mrad$ for
vertical coordinates.
The mass resolution of the $\pi^0$-meson registered with the gamma-quanta detector was obtained to be $\sim 15~MeV$ (see Fig. \ref{pi0}).

\section {The SVD-2 physics program.}

The initial goals of the experiment were the following:
\begin{enumerate}
\item Measurement of the total charm production cross section on $Si$, $C$ and $Pb$ targets at the threshold energy and the study of the cross section $A$-dependence.
\item Measurement of the differential $x_F$ and $p_T$ spectra and study of the leading effect of the charmed particles.
\item Investigation of the possible influence of the "intrinsic charm" in the proton on the inclusive charm spectra.
\end {enumerate}

\begin{center}
\begin{figure}[ht]
\centering
\includegraphics[width=0.45\textwidth]{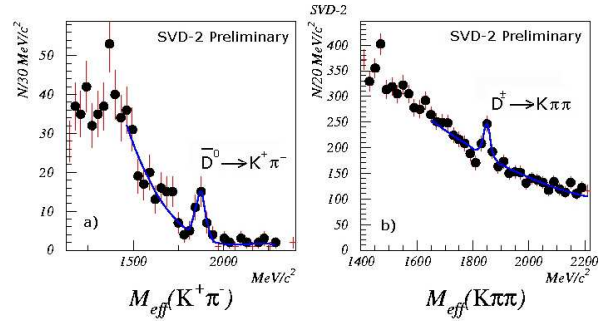}
\caption{a) The $K^+ \pi^-$ invariant mass spectrum. b) Combined $K^+ \pi^- \pi^-$ and $K^- \pi^+ \pi^+$ invariant mass spectrum.}
\label{d0}
\end{figure}
\end{center}

Since the cross section of the charm production near threshold energy is small, the complicated algorithms for the effective selection of the events with charmed particles are needed. This work is in progress now and we hope to obtain a few hundred events with $D$-mesons by the end of 2005. The Fig. \ref{d0}a illustrates invariant mass distribution for ($K^+\pi^-$)-system with both tracks having large impact parameters in x,y planes and decay point within 3 mm from a primary one. The Fig. \ref{d0}b shows the combined $K^+ \pi^- \pi^-$ and $K^- \pi^+ \pi^+$ and the charged $D$-meson signal is visible in this spectrum.

\begin{figure}[ht]
\centering
\includegraphics[width=0.5\textwidth]{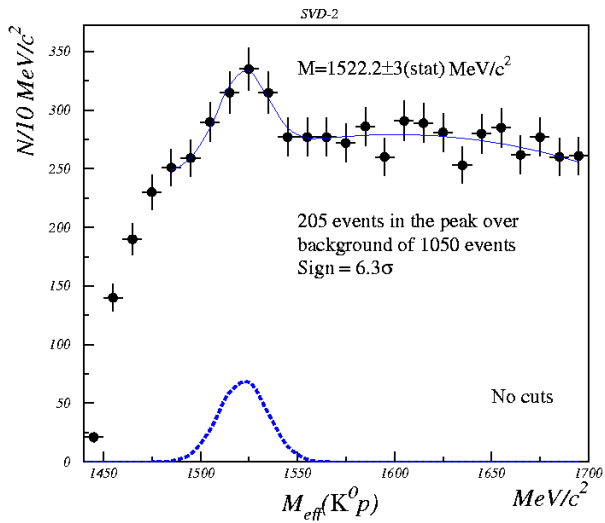}
\caption{The $(pK^0_s)$ invariant mass spectrum for $K^0_s$ decaying inside the vertex detector.}
\label{theta1}
\end{figure}

\begin{figure}[ht]
\centering
\includegraphics[width=0.51\textwidth]{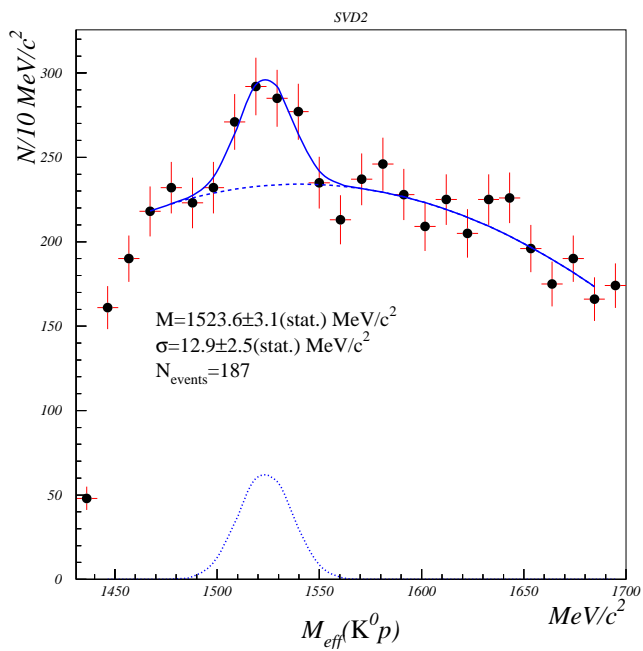}
\caption{The $(pK^0_s)$ invariant mass spectrum for $K^0_s$ decaying outside the vertex detector.}
\label{theta2}
\end{figure}

In the fall of 2003 the SVD collaboration started searches for the exotic $\Theta^+$-baryon. This exotic resonance was predicted by Diakonov, Petrov and Polyakov\cite{dpp} and by that time was observed by LEPS\cite{leps}, DIANA\cite{diana} and CLAS\cite{clas} collaborations.
The clean sample of $K^0_s$-mesons detected in vertex detector allowed
us to search for the decay $\Theta^+ \rightarrow pK^0_s$ and the narrow peak was observed in the ($pK^0_s$) invariant mass spectrum\cite{theta1}.
During the 2004-2005 SVD collaboration continued the study of ($pK^0_s$)-system using improved algorhitms of tracking in the vertex detector and also using another sample of $K^0_s$-mesons decayed after the vertex detector. So, two different samples of $K^0_s$, statistically independent and belonging to different phase space regions were used in the analyses and a narrow baryon resonance with the mass $M=1523\pm 2(stat.)\pm 3(syst.)\ MeV/c^2$ was observed in both samples of the data (see Fig. \ref{theta1} and \ref{theta2}). The statistical significance was estimated to be of $8.0~\sigma$ (392 signal over 1990 background events). Using the part of events reconstructed with better accuracy the width of resonance was estimated to be $\Gamma < 14~MeV/c^2$ at 95\% C.L.\cite{theta2}

\section {The current status of the SVD-2M detector.}

In the 2003 the a experiment using SVD-2 detector was proposed\cite{therm1,therm2}. The goal of the proposed  "Thermalization" project is to investigate the collective behaviour of particles in the process of high multiparticle production in $pp$ (or $pN$) interactions:

$pp \rightarrow n_{\pi} + 2N$

\noindent
at the proton energy $E_p = 70\ GeV$. At present the multiplicity distribution at this energy is measured up to the number of charged particles $n_{ch} = 20$\cite{mult8}. The corresponding scaling variable $z = n_{ch}/\bar{n}_{ch} = 3.5$. The kinematics limit is $n_{\pi,th} = 69$, $z_{th} = 8.2$. Here, $n_{\pi,th}$ is the maximal number of charged and neutral pions allowed by energy-momentum conservation. In the frame of the "Thermalization" project the study of the events with multiplicity $n_{\pi} = 20-35$, $z = 3-5$ is planned. For the purposes of this upcoming experiment the SVD-2 setup is now being upgraded.

The modernized setup (SVD-2M) will include:

\begin{enumerate}
\item A liquid hydrogen target with diameter of 2 cm and length of 2.5 cm.
\item A new vertex detector with a higher efficiency and a lower multiple scattering.
\item A new tracking system based on 9 double planes of straw tubes with a resolution of 200 $\mu m$. This system will replace the first set of proportional chambers. With that the detection of charged particles will be noticeably improved.
\item New fast electronics for the gamma-quanta detector readout.

\end{enumerate}

It is also planned to increase intensity capabilities of proportional chambers by installing beam killers and to increase the efficiency of Cerenkov detector up to 90\%. The "Thermalization" experiment needs special trigger on the high multiplicity events and such a trigger based on microstrip silicon planes is also being created.

With the SVD-2M setup it is planned to continue study of charmed particles production (with $Si$, $C$ and $Pb$ targets) and the searches for the new hadron states (using hydrogen and nuclear targets). The program is planned for years 2005-2007 with obtaining a total luminosity of $\sim 50~nb^{-1}$.

Authors are grateful to the Organizing Committee for the support and hospitality.

\end{document}

%% file: authors.tex
\def\groupsinp{\affiliation{D.V. Skobeltsyn Institute of Nuclear Physics, Lomonosov Moscow State University, 1/2 Leninskie gory, Moscow, 119992 Russia}}
\def\groupihep{\affiliation{Institute for High-Energy Physics, Protvino, Moscow oblast,  142284, Russia}}
\def\groupjinr{\affiliation{Joint Institute for Nuclear Research, Dubna, Moscow oblast, 141980, Russia}}

\groupsinp
\groupihep
\groupjinr

\author{P.~Ermolov} \groupsinp
\author{A.~Kiriakov} \groupihep
\author{A.~Kubarovsky} \groupsinp
\author{V.~Nikitin} \groupjinr
\author{V.~Popov} \groupsinp
\author{I.~Rufanov} \groupjinr
\author{L.~Tikhonova} \groupsinp
\author{V.~Volkov} \groupsinp
\author{A.~Vorobiev} \groupihep

\collaboration{on behalf of the SVD Collaboration} \noaffiliation